# Suspect AI: Vibraimage, Emotion Recognition Technology, and Algorithmic Opacity

**Author**: James Wright

**Affiliation**: The Alan Turing Institute

**Abstract**

Vibraimage is a digital system that quantifies a subject's mental and emotional state by analysing video footage of the movements of their head. Vibraimage is used by police, nuclear power station operators, airport security and psychiatrists in Russia, China, Japan and South Korea, and has been deployed at an Olympic Games, FIFA World Cup, and G7 Summit. Yet there is no reliable evidence that the technology is actually effective; indeed, many claims made about its effects seem unprovable. What exactly does vibraimage measure, and how has it acquired the power to penetrate the highest profile and most sensitive security infrastructure across Russia and Asia?

I first trace the development of the emotion recognition industry, before examining attempts by vibraimage's developers and affiliates scientifically to legitimate the technology, concluding that the disciplining power and corporate value of vibraimage is generated through its very opacity, in contrast to increasing demands across the social sciences for transparency. I propose the term 'suspect AI' to describe the growing number of systems like vibraimage that algorithmically classify suspects / non-suspects, yet are themselves deeply suspect. Popularising this term may help resist such technologies' reductivist approaches to 'reading' – and exerting authority over – emotion, intentionality and agency.

**Funding**

This work was supported by Michelin, the Fondation France-Japon de l'EHESS, and the UK Economic and Social Research Council for the *Sustainable Care: Connecting People and Systems* programme [ES/P009255/1], 2017-21, Principal Investigator Sue Yeandle, University of Sheffield.

**Acknowledgements**

The author gratefully acknowledges the assistance of members of the ELSYS Japan team, who generously gave up their time to meet me in Tokyo and answer questions about vibraimage. I would also like to thank Hallam Stevens for providing invaluable feedback on a draft version of this paper. Russell Henshaw and Arthur Thompson also provided extremely helpful comments and suggestions.

**Introduction**

As I sat in the meeting room of a nondescript office building in Tokyo, the senior managing director of a company called ELSYS Japan discussed my emotional and psychological state, referring to a series of charts and tables displayed on a large screen at the front of the room:

> Aggression (*kōgekisei*)… 20-50 is the normal range, but you scored 52.4… this is a bit too high. Probably you yourself didn't know this, but you're a very aggressive person, potentially… Next is stress (*sutoresu*). Your stress is 29.2, within the range of 20-40, with a statistical deviation of 14 – that's OK… I think you have very good stress... Just tension (*kinchō*) – your [average] value is within the range, but because your statistical deviation is high – over 20 – so you're a little tense. Mental balance (*seishin no baransu*) is 64 from a range of 50-100, so it fits correctly in the range… Charm (*miryoku*)… 74.6 is pretty good. Now, neuroticism is 35.3, this is also in the range, but the statistical deviation is high. But some people have a high score the first time they are measured. There are people who have high scores for neuroticism as well as for tension, yes. People who possess a delicate heart.
>
> (Interview, 17 April 2019[i])

The director's seemingly authoritative statements were based on an assessment of various measurements produced by 'vibraimage', a patented[ii] system developed to quantify a subject's mental and emotional state through an automated analysis of video footage of the physical movements of their face and head. This system, which is distributed in Japan by ELSYS Japan under the product brands 'Mental Checker' and 'Defender-X', provides numerical values for levels of: aggression; tension; balance; energy; inhibition; stress;

suspiciousness[iii]; charm; self-regulation; neuroticism; extroversion; and stability, categorising these automatically into positive and negative 'emotions'. Mental Checker generates an impressive array of statistical data neatly arranged across tables, a pie chart, histogram and line chart, which produces an image of mathematical precision and solid scientific legitimacy (see Figure 1). The report also provides a visualisation of what ELSYS Japan terms an 'aura'[iv] – a horizontal colour-coded bar chart indicating the frequency of micro-vibrations of a subject's head, superimposed against a still image of their face.

Vibraimage technology has already entered the global security marketplace. It was deployed at the 2014 Sochi Olympics (Herszenhorn, 2014), the 2018 PyeongChang Winter Olympics, the 2018 FIFA World Cup in Russia, and at major Russian airports, to detect suspect individuals among crowds (JETRO, 2019). It has also been used at Rosatom, the Russian State Atomic Energy Corporation, in experiments focusing on monitoring the professionalism of dozens of workers operating the handling and disposal of spent nuclear fuel and radioactive waste (Bobrov et al., 2019, Shchelkanova et al., 2019), as well as diagnosing their psychosomatic illnesses (Novikova et al., 2019). In Japan at time of writing, Mental Checker and Defender-X are being used by one of the largest technology and electronics companies, NEC[v], to vet staff at nuclear power stations, and by one of the leading security services companies, ALSOK, to detect and potentially deny entry to or detain suspicious individuals at major events including the G7 Summit in 2016, as well as at sporting events and theme parks across Japan (Interview with managers at ELSYS Japan, 17 April 2019). Sales managers at ELSYS Japan expected that the technology would be used at the 2020 Tokyo Olympics (Nonaka 2018, p. 148, Interview with managers at ELSYS Japan, 17 April 2019), an event that has spurred a significant increase in spending on domestic security services and infrastructure, with an estimated market growth of 18 per cent between 2016 and 2019

(Teraoka, 2018).[vi] ELSYS Japan's customers also include Fujitsu and Toshiba, which have considered 'incorporat[ing] [vibraimage]… into their own recognition technologies to differentiate their original products' (Nonaka 2018, p. 147), and managers told me that Mental Checker has been used by an unspecified number of Japanese psychiatrists to confirm diagnoses of depression.

In South Korea, the Korean National Police Agency, Seoul Metropolitan Policy Agency and several universities have collaborated on research aiming to establish the use of vibraimage in a video-based 'contactless' lie detection system as an alternative to polygraph testing (Lee and Choi, 2018, Lee, Choi & Jang, 2018), while in China, it has been deployed in Inner Mongolia, Zhejiang, and other locations to identify suspects for questioning and detention and has now been officially certified for use by Chinese police forces (Choi, Kim & Hu, 2018, Choi et al., 2018).[vii] Various other corporate applications of vibraimage have also been proposed: a brochure produced by ELSYS Japan suggests using Mental Checker to find out how employees really feel about their company; measure their levels of stress, fatigue, and 'potential ability'; counter employees' accusations of bullying and abuses of power in the workplace; and even 'to know the risk of hiring persons who might commit a crime' (ELSYS Japan brochure, undated). The brochure provides a screenshot of a suggested employee report, with grades (A+, B-, C, etc) for various scores, including: stability; fulfilment and happiness; social skills; teamwork; communication; ability to take action; confidence; aggressiveness; stress tolerance; sense of responsibility; and ability to 'recognise reality' (*genjitsuninshiki*).[viii]

Vibraimage forms one part of the rapid growth in algorithmic security, surveillance, predictive policing and smart city infrastructure across urban East Asia, enabling the 'active sorting, identification, prioritization and tracking of bodies, behaviours and characteristics of subject populations on a continuous, real-time basis' (Graham & Wood 2003, p. 228). Amid an international boom in both surveillance technologies and AI systems designed to extract as much information as possible from digital photographic and video data relating to the body, companies are developing algorithms that move beyond facial recognition intended to identify individuals, and increasingly aim to analyse their behaviour and emotional states (AI Now Institute, 2018, pp. 50-52). The digital emotion recognition industry was worth up to $12bn in 2018, and continues to grow rapidly (Ibid.).

As the concepts of algorithmic regulation and governance (Goldstein, Dyson, & Nemani, 2013, Introna, 2016) are increasingly becoming a reality, not least through the catalysing effects of state responses to the COVID-19 pandemic in the form of digital contact tracing, transparency has become a key theme in critiques of black-boxed algorithms and AI, including those used in emotion recognition. This is particularly the case with machine learning, in which algorithms recursively adjust themselves in ways that can quickly become inexplicable even to data science experts. As Maclure puts it, 'we are delegating tasks and decisions that directly affect the rights, opportunities and wellbeing of humans to opaque systems which cannot explain and justify their outcomes' (2019, p. 3). Transparency is linked to and overlaps with values of comprehensibility, explicability, accountability of those developing and operating these algorithms, and social justice. It tends to be seen as a vital component of ethical or 'good' AI (Hayes, van de Poel & Steen, 2020, Floridi et al., 2018, Leslie, 2019).

Burrell describes three types of algorithmic opacity: '(1) opacity as intentional corporate or institutional self-protection and concealment and, along with it, the possibility for knowing deception; (2) opacity stemming from the current state of affairs where writing (and reading) code is a specialist skill and; (3) an opacity that stems from the mismatch between mathematical optimization in high-dimensionality characteristic of machine learning and the demands of human-scale reasoning and styles of semantic interpretation' (2016, pp. 1-2). Danaher (2016) likewise draws a distinction between opacity and hiddenness, where opacity refers to the incomprehensible or inaccessible ways in which systems work, and hiddenness refers to the covert and hidden manner in which data is collected and used. As I will show in this paper, both of the latter definitions apply in the case of vibraimage. Data used by the system may be collected and analysed in a hidden or covert manner (for example, via CCTV at an event or in a public place), and its exact method of analysing this data is also opaque – it is not clear to its subjects how the system works or what exactly it is quantifying.

This paper uses the case of vibraimage to examine issues around opacity and the work that it does for companies and governments in the provision of security services, by attempting to shed some light on the algorithms of vibraimage and on its imagined and actual uses, as far as possible based on publicly available data. What exactly does vibraimage measure, and how does the data the system produces, processed through an algorithmic black box, deliver reports that have acquired the power to penetrate the heart of corporate and public security systems and infrastructure involved in the highest profile and most sensitive security tasks in Russia, Japan, China, and elsewhere? The first part of the paper examines emotion detection techniques and their digitalisation. The second part focuses on vibraimage, and how its proponents, many of whom are involved in commercial relationships with companies distributing it, have engaged in processes of scientific legitimation of the technology as well as making claims for its actual and potential uses. The final section looks at how the disciplining power and corporate value of vibraimage is generated through its very opacity, in stark contrast to increasingly urgent demands across the social sciences and society more broadly for transparency to develop 'good AI'. I propose the term 'suspect AI' reflexively to describe the increasing number of algorithmic systems such as vibraimage, in operation globally across law enforcement and security services, which involve automatically classifying subjects as suspects or non-suspects. Popularising this term may be one way to resist such reductivist approaches to reading and exerting authority over human emotion, intentionality, behaviour, and agency.

**Emotion recognition based on facial expressions**

Psychologist Paul Ekman has pioneered research exploring the relationship between emotions and facial expressions since the 1960s, building on Charles Darwin's work on the evolutionary connections between the two among animals, including humans (Darwin, 2012 [1872]). Ekman conducted experiments aiming to categorise a number of basic emotions (such as anger, contempt, disgust, fear, happiness, sadness and surprise) in order to demonstrate that these are universal across all cultures and societies, and expressed in universal ways through similar facial expressions (Ekman, 1992). This work was highly influential because the experiments Ekman and his collaborators conducted around the world seemed to provide overwhelming empirical evidence that individuals of all cultures were able 'correctly' to categorise the expressions of people of their own and other cultures provided in photos, matching them to the 'basic emotions' that they were supposed to express (Ekman, 1971).

Ekman further argued that facial expressions could be used to identify incongruities between professed and 'real' emotions, enabling facial expression analysis to be used for lie detection (see, for example, Ekman & Friesen, 1969). This attracted substantial interest from corporations concerned with ensuring the honesty of employees or gaining covert insights in business negotiations, as well as from governments and security forces concerned with identifying dissimulating and suspect individuals. Ekman and collaborators in this field such as David Matsumoto have formed their own companies, running workshops and consulting with companies and public bodies on using techniques to read subjects' facial micro-expressions and behavioural cues to evaluate personality, truthfulness and potential danger. In 2001, Ekman was named by the American Psychological Association as one of the most influential psychologists of the twentieth century (APA, 2002).

The identification of emotions through facial expressions underwent digitalisation via machine learning techniques pioneered since the mid-1990s by Rosalind Picard and Rana el Kaliouby at the Massachusetts Institute of Technology. They commercialised this new field of 'affective computing' in the form of the venture capital-backed company Affectiva, founded in 2009, which provides emotional analysis software to businesses based on algorithms trained on its large databases of facial expressions (Johnson, 2019). According to Affectiva, this enables a test subject's emotional responses to, for example, TV commercials, to be tracked in real time. With the recent boom in facial recognition technology, emotion recognition now represents a rapidly developing area of AI development, and is in use across various industries including recruitment and marketing research (Devlin, 2020). A growing number of companies offer emotion recognition services based on the analysis of facial expressions, including the biggest tech corporations: Microsoft (Emotion API), Amazon (Rekognition), Apple (Emotient, for which Ekman acted as an advisor), and Google (Cloud

Vision API). In Japan, the robot 'Pepper', available on lease from telecoms and technology company SoftBank, incorporates emotion recognition software, and was marketed as the 'world's first humanoid robot that reads human emotions' (Wright, 2019).

Such systems are increasingly also being used in border protection and law enforcement to identify dissimulating and otherwise suspect individuals, regardless of substantial evidence or proof about efficacy. Since 2007, the Transportation Security Administration (TSA) spent $900 million on a 'behavior-detection program' entitled Screening Passengers by Observation Technique (SPOT), which was eventually ruled ineffective by both the Department of Homeland Security and the Government Accountability Office in 2013 (GAO, 2013). Ekman consulted on SPOT, and the system incorporated some of his techniques; his company has also provided consulting services to US courts (Fischer, 2013). Another system, Automated Virtual Agent for Truth Assessments in Real-Time (Avatar), was developed for lie detection aimed at migrants on the US-Mexico border (Daniels, 2018), while the European Union recently trialed the iBorderCtrl system, supplied by the consortium European Dynamics and funded by the EU's Horizon 2020 fund, to use the interpretation of micro-expressions to detect deceit among migrants in Hungary, Greece, and Latvia (Boffey, 2018).[ix]

Recently, much work on facial expression analysis for emotion recognition has come under increasing scrutiny. The most basic critique is that one does not necessarily smile when one is happy – common sense suggests that facial expressions do not always, or even often, map to inner feelings, that emotions are often fleeting or momentary, and that facial expressions and their meaning are highly dependent on social and cultural context. A major review of evidence relating to the link between emotions and facial expressions (Barrett et al., 2019) summarises these and other more sophisticated critiques. These include:

> 1. *Limited reliability* (i.e., instances of the same emotion category are neither reliably expressed through nor perceived from a common set of facial movements).
> 2. *Lack of specificity* (i.e., there is no unique mapping between a configuration of facial movements and instances of an emotion category).
> 3. *Limited generalizability* (i.e., the effects of context and culture have not been sufficiently documented and accounted for).
>
> <div align="right">(Barrett et al., 2019, p. 3; original emphases)</div>

Barrett et al. argue that approaches positing a limited number of prototypical basic emotions that can be 'read' through universal facial expressions fail to grasp what emotions are, and what facial expressions convey.

In anthropology, the 'affective turn' has drawn attention to a productive distinction between affect and emotion – the former a pre-cognitive sensory response or potential to affect and be affected, and the latter a more culturally-mediated expression of feeling. Daniel White describes this as the difference between 'how bodies feel and how subjects make sense of how they feel' (2017, p. 177). These nuances are overlooked in the field of emotion recognition, which reduces emotion to a highly reductive and simplistic model that is digitally scalable. Barrett argues that:

> emotion isn't a simple reflex or a bodily state that's hard-wired into our DNA, and it's certainly not universally expressed. It's a contingent act of perception that makes sense of the information coming in from the world around you, how your body is feeling in the moment, and everything you've ever been taught to understand as emotion. Culture to culture, person to person even, it's never quite the same.
>
> <div align="right">(Quoted in Fischer, 2013)</div>

Thus, we might define the process of interpreting one's own emotional state as one of making sense of an inner noise of biological signals and memories, in contextually contingent and socio-culturally mediated ways, and placing them into – and in the process co-constructing – socio-culturally mediated categories. It may also sometimes involve not definitively categorising or making sense of these affective feelings. As this paper will show, it is the very ambiguity or malleability of this process that may help make vibraimage a convincing technology of emotion recognition, and provide authority to its reports.

Given these growing critiques of Ekmanian theories of universal basic emotions expressed through facial expressions, researchers at the organisation AI Now have concluded that, by extension, the digital emotion detection industry is 'built on markedly shaky foundations… There remains little to no evidence that these new affect-recognition products have any scientific validity' (AI Now Institute, 2018, p. 50). Baesler similarly argues that the use of emotion detection software by the TSA is 'unconfirmed by peer-reviewed research and untested in the field' (2015, pp. 60-61), while holding significant potential for harm through misuse. In common with broader critiques of AI from critical algorithm studies (e.g. Eubanks, 2018, Lum & Isaac, 2016), the use of machine learning methods involved in emotion recognition systems has been criticised for racial bias based on the data sets on which they are trained (Rhue, 2018). Indeed, Ekman's work not only constructs ethnocentric emotional categories, but also racial subject categories, for example in his construction, with Matsumoto, of the Japanese and Caucasian Facial Expressions of Emotion (JACFEE) stimulus set of 56 photos showing emotional expressions of archetypal 'Japanese' and 'Caucasian' subjects (Biehl et al., 1997, humintell.com), which continues to be used in psychology experiments. For all of these reasons, the increasingly widespread application of this technology has raised growing ethical and civil liberties concerns.

**'Welcome to the vibraimage world!'[x]: attempts to construct scientific legitimacy**

While the majority of digitalised emotion recognition systems are premised on facial expression analysis in order to determine the combination and intensity of the handful of so-called basic emotions that a subject is expressing, alternative technologies for evaluating emotional or mental states from other facial or bodily physiological data have also been developed. These include analyses of gait, voice, or eye movements, and of combinations of physiological data, such as polygraph lie detector tests that combine measurements of blood pressure, heart rate, respiration rate and galvanic skin response (electrodermal activity associated with sweating). An alternative technological approach, 'vibraimage', has also emerged, pioneered by Russian biometrist Viktor Minkin and developed since around 2000.[xi] This technology forms the basis for Mental Checker and Defender-X, the software products offered by ELSYS Japan, which is the Japanese affiliate of ELSYS Corp, a Russian company founded by Minkin.[xii]

Vibraimage involves recording a short video of a subject in order to measure and analyse 'vibrations' of the face and head: nearly imperceptible and involuntary micro-movements caused by muscles and the circulatory system in the neck and head. These movements are partly related to the vestibular system, which includes parts of the inner ear responsible for maintaining balance and spatial orientation, or equilibrioception. Minkin observes that the vestibular system is linked to certain psychological disorders, and argues that its function is intimately connected to emotional and mental states, which he describes as the 'vestibulo-emotional reflex' (Minkin and Nikolaenko, 2008). As a result, according to Minkin, data about involuntary head movements can be measured and analysed to generate information about these mental and emotional states, as well as inferring other information, such as the personality type of the subject (Minkin and Nikolaenko, 2017a).[xiii] Physical balance and stability are directly equated to mental and emotional balance and stability. Minkin, like Ekman, refers back to Darwin's 1872 work on the biological link between facial expression and emotion in humans and animals. While Ekman developed this work by focusing on facial expressions, Minkin focuses on biological and mathematical links between muscular activity and brain activity, citing the work of nineteenth century Russian physiologist Ivan Sechenov (Minkin, 2017, p. 18) and various others who have contributed to the field of psychophysiology (a branch of psychology that explores links with physiology), including Sigmund Freud, Konrad Lorenz, and Ivan Pavlov, as well as Norbert Wiener's theory of cybernetics (Ibid., p. 6). Minkin refers to his theory variously as an 'applied theory of psychophysiology of motion', and 'the thermodynamic model of emotions' (Ibid., p. 56), because he draws a direct link between specific emotional-mental states, muscular activity that can be measured through micro-vibrations of the head, and the energy this muscular activity expends: 'physiological and psychophysiological processes proceeding in a human body are associated with the exchange of energy and information within or between human physiological systems' (Ibid., p. 50).[xiv] According to this theory, involuntary movement of the face and head is emotion, intention, and personality made visible.

Vibraimage technology also appears to have roots in cold weather experiments conducted in the Soviet Union. A senior manager at ELSYS Japan mentioned that the algorithms driving Mental Checker, which were used to make the connections between head vibration data and particular emotional and mental states, had been trained using a proprietary big data set. When I asked about this data set, she told me:

> It's actually from the Russian side [i.e. ELSYS Russia]… They have like data from actual – they put some people in cold weather and examine what kind of vibrations they will have under such circumstances. So some people are really frightened and they have their vibrations, and some have the actual examinations of like 100,000 people. So those data – even America, the US, cannot do such examinations because it's a human rights stuff. But in the Soviet Union, they could do that, so…
>
> [JW: So it's Soviet Union-era data?] Yes, that's why other companies cannot make such a system. Except in North Korea!
>
> (Interview, 17 April 2019[xv])

Minkin does not refer to this 100,000-piece data set from the Soviet Union in his publications on vibraimage, although it is mentioned in at least one Japanese newspaper report about ELSYS Japan (Saito, 2016), and Minkin worked in a state biometrics lab during the Soviet era (JETRO, 2019), which may feasibly have provided him with access to such data. In a subsequent email from ELSYS Japan, the same senior manager quoted above stated that 'we got the information directly from Mr. Viktor Minkin. The experiments have been done with people aged between 2 months to 90 years old with various nationalities. However, the detailed information has not been disclosed, so it's not referred in any of his work [sic]' (email, 26 February 2020). If this data exists and was used to train the vibraimage algorithms, this would raise serious questions about the ethics of using an AI system developed using data apparently obtained in a highly unethical manner. It would also raise technical questions about how exactly this data set was used to derive algorithmic connections between specific types and intensities of emotional-mental states and head movements – a significant piece of the puzzle of how vibraimage works that is conspicuously absent from publications aimed at establishing its scientific veracity. If the data does not in fact exist, it equally raises questions about why such a narrative of human experimentation is viewed as a valid and effective means by which to legitimise the technology.

On 20 January 2020, a search of Google Scholar for the term 'vibraimage' yielded 287 results.[xvi] Of these, a large proportion were written or co-authored by users with a strong commercial interest in the success of the technology, such as Minkin himself, or employees of ELSYS Russia or international companies affiliated with it that hold distribution rights to vibraimage technology, thus introducing significant potential for biased findings.[xvii] 41 of these papers were published in the proceedings of two conferences on vibraimage technology organised and hosted by ELSYS Russia. The first, entitled 'Modern Psychophysiology: The Vibraimage Technology', was held in St Petersburg in June 2018, with the apparent support and participation of the European Academy of Natural Sciences and the Russian Biometric Association among others. A second conference was held in June 2019, and a third was planned for 2020. Very few articles on vibraimage appear to have been published in academic journals with rigorous peer review processes; several appear in journals such as the Journal of Behavioral and Brain Science, and Intelligent Control and Automation, both published by Scientific Research Publishing (SCIRP), a company included in Jeffrey Beall's list of predatory or questionable academic open access publishers, with potentially poor journal standards.[xviii]

Many papers that feature vibraimage technology proceed from the a priori assumption that its reliability and effectiveness (and the existence of a 'vestibulo-emotional reflex' on which it is premised) has already been proven, and involve conducting vibraimage tests on subjects and interpreting the results – rather than questioning or attempting to prove or verify the efficacy of the underlying technology itself. In one article, Minkin and Yana Nikolaenko, chief psychologist at ELSYS Russia, state that they aim to 'introduce… a new term, vestibular emotional reflex or vestibular energy reflex (VER)' (Minkin and Nikolaenko, 2008, p. 196), but proceed to lay out technical details of vibraimage rather than providing any evidence that such a reflex exists, or about the nature of its connection to head movements. In another paper, Minkin and Nikolaenko use vibraimage analysis to group 'criminal' and 'non-criminal' research subjects according to personality types (Minkin and Nikolaenko, 2017b). The data itself appears to show a random distribution, but the authors interpret it to fit into a schema of personality types, although there seems no way to verify the accuracy of the way it assigns subjects to personality categories, particularly since their analysis is based on uncovering 'unconscious' responses. Nevertheless, they state that this methodology could be used to predict which personality types are prone to commit crimes:

> If we assume that the revealed picture of differences between conscious and unconscious responses accurately reflects the hidden information, then this method can be the basis for identifying individuals who intend to commit criminal acts or who are predisposed to commit such acts.
>
> (Minkin and Nikolaenko, 2017b, p. 459)

Similarly, a paper by Nikolaenko (2018) aims to measure the 'level of delinquency' among adolescents by identifying those with personalities more likely to commit crimes. While much of the predictive policing industry in the United States, dominated by companies such as PredPol, IBM, Palantir and HunchLabs, relies on applying analytical techniques such as machine learning to big data sets and searching for patterns 'to identify likely targets for police intervention and prevent crime or solve past crimes by making statistical predictions'

(Perry et al., 2013, p. xiii; cf. Lum & Isaac, 2016)[xix], vibraimage promises to identify suspect criminal types through a twenty-first century version of phrenology.[xx] Although in their article mentioned above, Minkin and Nikolaenko provide the caveat that, '[t]he question of estimation accuracy of psychophysiological parameters of an individual, certainly, demands a larger set of statistics, and it was not investigated in this paper' (Minkin and Nikolaenko, 2017b, p. 459), it is unclear how any accuracy of grouping by personality type presented in such papers could in fact be accomplished, since the results appear to be neither falsifiable nor reproducible.

Very few peer-reviewed academic articles outside ELSYS Russia's vibraimage research ecosystem seem to have been published. However, one such article by Japanese researchers with no apparent commercial connection to ELSYS Russia or ELSYS Japan attempted to establish the efficacy of vibraimage in measuring actual or latent mental states in order to identify suspect individuals, by comparing various established paper-based psychological methodologies with vibraimage. The authors found almost no statistically significant correlations between the results of existing psychological tests and those produced by vibraimage, leading them to the carefully worded conclusion that:

> present psychological measurement research cannot identify what is being measured by the indicators that express mental state based on vibraimage technology… Since we do not know what the parameters are measuring, we cannot authoritatively state that it is not effective as a system to detect suspicious people. What we can say as a result of this research is that we do not understand what it is measuring. We cannot say that the possibility of detecting a suspect is zero.
>
> (Ōkubo et al., 2018, p. 25; author's translation)

It is important to note the importance of this opacity – not knowing what exactly vibraimage is measuring, and what these measurements mean – which I will discuss further below. The only other paper directly addressing the question of the level of accuracy of vibraimage was written by Minkin himself (2019). Observing that alternative 'standardized measures (standards) for measuring the psychophysiological state (PPS) do not currently exist' (Ibid., p. 212), Minkin compares the accuracy of head movement measurements provided by different versions of the same vibraimage software. Unsurprisingly, he concludes that vibraimage has a very low measurement error rate.

Minkin credits the inspiration for his 'thermodynamic model of emotion' to Libb Thims and Georgi Gladyshev (Minkin, 2017, p. 17). This model is proposed in an article co-authored by Minkin, Gladyshev and Thims, and published on ELSYS Russia's homepage. This article, referenced by many contributors to the 2018 and 2019 conferences on vibraimage technology as evidence for the efficacy of the technology, argues that: 'head balance for person without consciously movements could be considered as isolated thermodynamic system and any internal energy as emotion would change the balance of this internal system and realized by movements or vibrations [sic]' (Minkin, Gladyshev and Thims, undated).

Gladyshev is a Russian scientist who has theorised the role of 'hierarchical thermodynamics' in living systems – an unorthodox and disputed area of physics. Thims is a self-described 'American electrochemical engineer, thermodynamicist, and physicochemical free thinker'[xxi] who appears to operate a series of websites including a Wikipedia-like website focusing on his interests in 'human thermodynamics', atheism, and historical figures with high IQ scores, as well as a separate website wholly dedicated to human thermodynamics. Across the content of both of these websites, Thims argues in a somewhat convoluted fashion that humans are molecules, and that social interactions are chemical reactions; in an extreme case of reductivism, this claim is presented not as a metaphor but as a literal scientific fact. Thims also ran his own self-published journal entitled 'Journal of Human Thermodynamics'[xxii] between 2005 and 2016, to which Minkin and Gladyshev have contributed articles, and Gladyshev has acted as a peer reviewer,[xxiii] although the Wikipedia-like peer review process described on the website does not meet the standards of most academic journals[xxiv], and Thims' own articles include titles such as 'Thermodynamic Proof that Good Always Triumphs over Evil.'[xxv] One of Thims' websites includes several pages, which are recorded in the pages' update information as having been created by Thims himself, documenting and refuting the critiques of various detractors, most of whom have described his work as pseudoscience[xxvi]; this catalogue of critics includes Gladyshev himself, who writes to Thims: 'I believe that you have created an incredible mess. I'm beginning to understand that you do not know "what science is?"'[xxvii]

My purpose in discussing Thims' websites is not to impugn the airing of his philosophy of human thermodynamics via personal websites, but rather to throw light on the intellectual origins and theories underpinning vibraimage. Minkin presents a similar type of reductivist philosophy to Thims, and is clearly inspired by Thims' views in seeing human behaviour as interpretable through the measurement of psychophysiological data, while specifically disregarding all other contextual information about a subject. For example, in one of Minkin and Nikolaenko's experiments to measure intelligence type (2017b), subjects divided into three groups according to levels of 'deviant behavior' (labelled as 'drinkers', 'criminals', and 'lawyers') take one kind of written test, while vibraimage measures their 'unconscious' response to the same questions, with the differences between the 'conscious' and 'unconscious' answers proving the value of vibraimage in uncovering the underlying truth about these subjects. In this way, the authoritativeness of vibraimage is built on actively undermining the agency of the subject in consciously determining their own mental-emotional state.

By measuring vibrations of the head, Minkin argues, all kinds of information about a person can be inferred: the characteristics of mental-emotional states listed above, but also character traits including truthfulness, loyalty, human potential, personality and intelligence types, 'deviant behavior' and delinquency, and the propensity to commit crimes in the future. Perhaps most troubling is Minkin and Nikolaenko's proposed application of vibraimage in *1984*-style techno-fascist tests of loyalty to a company or nation:

> Programs of the loyalty analysis to the principles and values of any state, society or production company can be based on the suggested method. Passing such a test of loyalty or human variability… may be in the future an integral part of obtaining a visa, along with scanning fingerprints. Not so long ago it seemed that biometric identification of a person is a pure fiction, but the next step in improving security will inevitably be the biometric identification of a psychophysiological state. Loyalty to the principles and values of the state is the same characteristic of a person as fingerprints, and its biometric identification is a very specific technical problem that has a unique solution, for example, using the vibraimage technology.
>
> (Minkin and Nikolaenko, 2017a, p. 129)

Since Minkin and others propose that vibraimage can be used not only to determine current mental-emotion states and personality types, but also to predict potential future behavior, these claims are, of course, difficult to verify.

**Conclusion: opacity and suspect AI**

Vibraimage is a system that appears to measure and quantify micro-movements of a subject's head from a video source, and convert these numbers into a variety of highly precise values that can be used to describe and categorise the mental-emotional states of subjects. Its algorithms provide measures to sort subjects, for example, into non-suspects and suspects: the latter people who have not necessarily (yet) committed any crime, offence, or otherwise negatively perceived action, but have failed a test of head movements and may therefore be subjected to detention, questioning, having their job, promotion or visa application rejected, or other pre-emptive disciplinary actions. As Graham and Wood note, because automated systems aim for exclusionary goals, '[a]lgorithmic systems thus have a strong potential to fix identities as deviant and criminal – what Norris calls the technological mediation of suspicion' (Graham & Wood, 2003, p. 234, referring to Norris, 2002). Yet as I have shown, the exact meaning and significance of what vibraimage measures is unclear: there is no reliable data pertaining to the accuracy of any parameter produced by vibraimage, and the process of quantification and categorisation of measurements into different values is opaque – a black-box process typical of many commercial algorithmic systems. Even if the way the algorithms worked were 'explicable', and even if the epistemological claims of Minkin and his collaborators concerning a 'vestibulo-emotional reflex' were credible, there is no coherent explanation of why certain intensities of head movements equate to a particular precise combination of emotions, behaviour, intent, or character. Vibraimage is a system about whose workings and efficacy complete transparency and knowledge seem not only unforthcoming but practically impossible to attain.

Nevertheless, the vibraimage developer, distributor, and user community is no longer simply a peripheral AI subculture – it is going mainstream, gaining momentum as an accepted form of security technology aligned with, and increasingly integrated into, facial recognition systems, and is already being used by major multinational corporations, nuclear facilities, and police forces across Russia and Asia. Facial recognition is already an everyday urban technology in China, and increasingly also in South Korea and Japan, where there are plans greatly to scale up its use across society, in cashierless shops and ATMs, at security gates at public and private events, at airplane and cruise ship boarding terminals, in job interviews, and in law enforcement (Gershgorn, 2020). Emotion recognition systems are increasingly included by default in commercial facial recognition systems, as in the case of Amazon's Rekognition and Google Cloud Vision API (McStay, undated). There exists the distinct possibility that vibraimage's algorithms will be – or have already been – incorporated into other facial recognition systems, such as those of ELSYS Japan's client NEC, one of the world's largest suppliers of facial recognition technology. Vibraimage may then be combined with other algorithms used in such systems, and rebranded, becoming part of yet more complex, opaque, algorithmic surveillance infrastructure, as suggested by the ELSYS Japan manager quoted in the introduction to this paper (Nonaka, 2018, p. 147).

As people come under increasing digital surveillance in public and commercial spaces, the apparent possibilities of harvesting additional biometric data from the body relating to mental-emotional states are clearly becoming an ever more tempting and important market for governments and corporations – particularly to identify potential non-compliance ('suspiciousness') among citizens, consumers, and migrants entangled in the kinds of increasingly complex and expensive socio-technical surveillance assemblages listed above that are becoming inescapable parts of daily life, movement, and work. These are areas in which a lack of state regulation of digital emotion recognition technology – especially relating to its application in geographical and socio-economic peripheries such as border enforcement, or more covert industries such as security services, law enforcement, or nuclear energy – may facilitate the introduction and propagation of systems such as vibraimage. Vibraimage in particular bestows even more power than a mainstream digital surveillance system would, because the results are at the same time precise yet somewhat ambiguous, open to interpretation, and cannot be validated.

Indeed, vibraimage wrests the power to interpret one's own feelings and intentions from the individual to the technology's operator. If emotion is a process of making sense of how one feels (White, 2017, p. 177) – making sense of an inner noise of biological signals and memories, in socio-culturally mediated ways, and placing these feelings into categories – vibraimage similarly involves the interpretation and categorisation of a kind of 'noise': raw data on head movement. ELSYS Japan's brochure on Mental Checker states that: 'Objective and stable data about the person can be obtained by Mental-Checker because it can visualise (quantify) the unconscious parts of the person neither by the opinions of the surrounding people nor by self-assessment [sic]' (ELSYS Japan brochure, undated). As suggested by the director's quote at the start of this article ('Probably you yourself didn't know this, but you're a very aggressive person, potentially'), proponents of vibraimage claim that it possesses the algorithmic authority to empower the operator to know subjects better than subjects know themselves, by directly accessing and revealing their unconscious. Making vibraimage's interpretation and knowledge of a subject's mental-emotional state more convincing and authoritative than that of the individual is accomplished through processes of quasi-scientific legitimation (articles, conferences), design (sleek data visualisations in the form of reports full of charts and histograms), and performance (video measurement procedure, appeals to the precision of AI), combined with the power of corporate state and private security infrastructure, and a lack of knowledge, awareness, or will to intervene in a booming market among politicians and regulators.

In one passage of their self-published book, *Vibraimage*, Minkin and Nikolaenko refer to Pavlov's famous series of experiments on dogs, in which Pavlov was able to condition a physiological response (salivation) to an external stimulus (the sound of a metronome or bell). They invert Pavlov's findings, suggesting that physiological responses can be quantified, analysed, and traced back to reveal the motive mental-emotional state that precipitated them. As Minkin and Nikolaenko put it, 'Now, practically, every person can act as a researched dog, and as Academician Pavlov since all that is needed for psychophysiological experiments is a computer and a web camera' (2017a, p. 129). The authors seem to suggest that use of vibraimage will democratise psychological insight: understanding what makes us tick – or salivate. But we might understand this analogy and power dynamic in a different way. The operator of vibraimage, whether corporation or government, is Pavlov, and the subject – employee or citizen – is the dog. Through this technological process, the subject is, at a profound level, made 'legible' (Scott, 1998) to governments and corporations. By this I do not mean that the subject is rendered transparently readable, but rather, through the power provided by the ambiguity and opacity of the system's algorithmic knowledge production, that the operator can *determine* – in both senses simultaneously – the emotions, character, current intentions, and future behaviour of the subject.

Vibraimage is a post-truth technology par excellence, in which the very opacity of the technology *is* the main characteristic and value of the technology. Ambiguity and uncertainty are leveraged, particularly in the performative interpretation of precise measures of imprecise subjective states: the black-boxedness of the technology corresponds to that of the human emotions that it claims to measure. Opacity and ambiguity constitute a currency that provides part of the system's authority and enables control to be exerted by the operator, introducing a highly unequal power relations and the potential for unjust decision-making and outcomes for those subjected to this system.

Ananny and Crawford have critiqued the idea that transparency alone is a sufficient condition for holding algorithmic assemblages accountable – 'that knowing is possible by seeing' (2018, p. 977). They argue that transparency as an ideal has numerous limitations and may even be harmful under certain conditions, and instead call for a model of algorithmic accountability that accepts these limitations and focuses on the sociotechnical assemblage that includes the algorithm as one element among many:

> if a system must be seen to be understood and held accountable, the kind of 'seeing' that an actor-network theory of truth requires does not entail looking inside anything – but across a system. Not only is transparency a limited way of knowing systems, but it cannot be used to explain – much less govern – a distributed set of human and non-human actors whose significance lies not internally but relationally.
>
> (Ibid., pp. 983-4)

Does it matter that we cannot see exactly how vibraimage's algorithms work, in order to hold accountable the broader surveillance assemblages of which they are a part? Indeed, even if we had full visibility of them, how would we know whether they actually 'work' in interpreting subjective mental and emotional states? Clearly, as Ananny and Crawford argue, transparency is not always the remedy for opacity, although scrutinising what we can of the algorithms themselves and scientific claims made about their effects is an essential part of critically analysing them, as is examining their position within broader power structures – seeing across the system of which they are one element. This is particularly the case in countries such as Russia and China – or even Japan and South Korea – with relatively little public accountability or critical media scrutiny of government technology policy or corporate technology strategy. Recognising the role of powerful states and corporations with little regulation in facilitating algorithmic assemblages such as vibraimage surveillance systems is essential in understanding not only the limitations of transparency as an ideal, as Ananny and Crawford point out, but also the limitations of the current push, primarily in Euro-American academic discourse, for an ideal ethics of AI that may have little practical relevance in other socio-cultural contexts.

The development and processes of scientific legitimation of vibraimage presented briefly in this paper constitute a case study of what I term 'suspect AI'. Vibraimage and several other AI emotion detection systems purport to identify suspects – persons suspected of an offence or some kind of ill intent, without proof or clear evidence. The adjective 'suspect' can be defined as 'not to be relied on or trusted; possibly dangerous or false'[xxviii], and I argue that this definition perfectly describes vibraimage itself – an AI system that is opaque, secretive, unproven, and dangerous – having the potential for significant harm. In situations in which full transparency and knowledge about the workings and effects of an algorithmic system are practically impossible to attain, popularising the term suspect AI, and adopting an analytical position of critical suspicion, may provide a form of activism and resistance, by interrogating such systems' constructed yet unproven legitimacy as they attempt to exert ever-greater authority over subjects.

[Figure 1]

Figure 1: Mental Checker report of author (photograph: author). Reproduced with permission from ELSYS Japan.

Yang, Y. (2017). China seeks glimpse of citizens' future with crime-predicting AI. *Financial Times*, July 23. Retrieved from https://www.ft.com/content/5ec7093c-6e06-11e7-b9c7-15af748b60d0

---

[i] Unless otherwise stated, quotes from this interview are translated from Japanese by the author.

[ii] Vibraimage is patented in Russia and the US, with a patent pending in Japan.

[iii] The measure for 'suspect' is a composite value of what the company describes as three 'negative' emotions (tension, aggression, and stress).

[iv] https://web.archive.org/web/20200206142847/http://www.psymaker.com/vibraimage/aura/.

[v] NEC is also one of the largest suppliers of facial recognition and other biometrics systems worldwide, particularly for public security. Among its myriad global contracts, it has a deal to implement facial recognition services to verify staff and athletes at the 2020 Tokyo Olympics (Gershgorn, 2020).

[vi] A senior official from the Tokyo Olympic committee was quoted in 2018 as stating: 'We're developing security as a national policy. I think it will be a success if we can make this a sales pitch for our country's latest security technology. That, too, is a sub-theme [of hosting the Games]' (Kanamori, 2019). Olympic Games have been a key market for security service companies; a senior manager from ALSOK in charge of preparations for security for Tokyo 2020 stated: 'The security industry was born as a legacy of the 1964 Games' (Tokyo 2020, 2018). ALSOK has already confirmed that during the (now postponed) Tokyo Olympics it plans to use facial recognition software together with a robot called 'Perseusbot' that can detect aggressive people among crowds, although it is unclear whether this system also utilises vibraimage technology (Kanamori, 2019, Japan Times, 2018).

[vii] Emotion recognition systems appear also to have been deployed in Xinjiang and more widely across different public security bureaux, although it is unclear whether these systems include vibraimage (Wong & Liu, 2019).

[viii] This can also be seen on page 10 of the following document: http://bit.ly/2v61Qsz.

[ix] For other applications of affect recognition to crime detection, see AI Now Institute, 2018, pp. 50-52.

[x] Quotation from ELSYS Russia slideshow, http://www.myshared.ru/slide/1035186/.

[xi] https://web.archive.org/save/http://www.elsys.ru/vibraimage_e.php.

[xii] To avoid confusion with ELSYS Japan and other international ELSYS affiliate companies, I will refer to the Russian company as ELSYS Russia.

[xiii] Minkin, as well as managers at ELSYS Japan, tends to conflate 'emotional state' and 'mental state'.

[xiv] An understanding of humans primarily in thermodynamic terms recalls the work of anthropologist Leslie White (2007 [1959]), who proposed a generalised theory of cultural evolution based on energy expenditure, which involved viewing biological organisms including humans as thermodynamic systems. White did not specifically focus on the expenditure of energy through human emotion at the individual physiological level, and his work is not referenced by Minkin or his collaborators.

[xv] This part of the interview with ELSYS Japan employees was conducted in English.

[xvi] I did not search for articles in other languages such as Russian, Korean or Japanese – although several results were listed in these languages, as 'vibraimage' was written as a keyword in the Roman alphabet.

[xvii] These include Minkin's close collaborators and employees listed on ELSYS Russia's website (https://web.archive.org/save/http://www.ELSYS.ru/addresses_e.php) as well as Kwan Choi, president of the South Korean distributor of vibraimage technology, and Yamauchi Hidetoshi (CEO) and Nonaka Kotoha, (manager) of ELSYS Japan.

[xviii] See https://beallslist.net.

[xix] In China, an AI system distributed by facial recognition company Cloud Walk aggregates and analyses data on subjects' movements and behaviour to identify suspicious individuals who may commit crimes (Yang, 2017). In Japan, the Kyoto Prefectural Police Department introduced a Predictive Crime Defense System in 2016. However, this system does not use machine learning partly because of the relative lack of data points – Japan's crime rate is very low. Instead, it uses statistical data about past crimes to try to identify efficient patrol routes for police officers (AIR, 2018, pp. 87-90).

[xx] The application of vibraimage for the detection of potential criminals was investigated by Zhelgang Police University in China from 2016-17, particularly focusing on the 'problem' areas of Tibet and Inner Mongolia. The technology was used directly to detain suspicious individuals, and the system has now been certified by Chinese police (Choi, Kim & Hu, 2018, p. 193). While evidence presented in Choi, Kim & Hu's paper points only to the apparent successes of the trial implementation, they do not provide detailed data, including false positive and false negative results.

[xxi] https://web.archive.org/web/20200206144835/http://web.archive.org/screenshot/https://twitter.com/libbthims?lang=en.

[xxii] https://web.archive.org/web/20200206144945/http://www.humanthermodynamics.com/Journal.html

[xxiii] See, for example: https://web.archive.org/web/20200206145046/http://www.eoht.info/page/SLTEP.

[xxiv] See https://web.archive.org/web/20200206145206/http://www.eoht.info/page/peer+review.

[xxv] *Journal of Human Thermodynamics* 2011, 7: 1-3.

[xxvi] https://web.archive.org/web/20200206145845/http://www.eoht.info/page/Libb+Thims+%28derogation%29.

[xxvii] https://web.archive.org/web/20200206150117/http://www.eoht.info/page/Libb+Thims+%28attack%29.

[xxviii] Oxford Dictionaries definition (see http://english.oxforddictionaries.com/suspect).